\documentclass[letterpaper,twocolumn,aps,prl,floatfix,superscriptaddress,noeprint]{revtex4-1}

\usepackage{float}	
\usepackage[caption=false]{subfig}
\usepackage{graphicx}	
\usepackage{mathtools}
\usepackage{physics}
\usepackage[colorlinks=true]{hyperref}
\usepackage[capitalise]{cleveref}

\graphicspath{ {./figs/}}

\newcommand{\Uzz}{U_{\rm II}}
\newcommand{\Uhar}{U_{\rm I}}

\newcommand{\Sc}{\mathcal{S}}
\newcommand{\Oc}{\mathcal{O}}
\newcommand{\Cc}{\mathcal{C}}

\begin{document}	
\title{Minimal Model for Fast Scrambling}
\date{\today}
\author{Ron Belyansky}
\email{rbelyans@umd.edu}
\author{Przemyslaw Bienias}
\author{Yaroslav A. Kharkov}
\author{Alexey V. Gorshkov}
\author{Brian Swingle}
\affiliation{Joint Center for Quantum Information and Computer Science, NIST/University of Maryland, College Park, Maryland 20742 USA}
\affiliation{Joint Quantum Institute, NIST/University of Maryland, College Park, Maryland 20742 USA}

\begin{abstract}
	We study quantum information scrambling in spin models with both long-range all-to-all  and short-range interactions.
	We argue that a simple global, spatially homogeneous interaction together with local chaotic dynamics is sufficient to give rise to fast scrambling, which describes the spread of quantum information over the entire system in a time that is logarithmic in the system size.
	This is illustrated in two tractable models: (1) a random circuit with Haar random local unitaries and a global interaction and (2) a classical model of globally coupled non-linear oscillators. 
	We use exact numerics to provide further evidence by studying the time evolution of an out-of-time-order correlator and entanglement entropy in spin chains of intermediate sizes.
	Our results pave the way towards experimental investigations of fast scrambling and aspects of quantum gravity with quantum simulators.
\end{abstract}
\maketitle
\pagenumbering{arabic}
{\it Introduction.---}%
The study of quantum information scrambling has recently attracted significant attention due to its relation to quantum chaos and thermalization of isolated many-body systems \cite{Deutsch1991,Srednicki1994,Rigol2008} as well as the dynamics of black holes \cite{Hayden2007,Lashkari2013,Shenker2014,Maldacena2016}.
Scrambling refers to the spread of an initially local quantum information over the many-body degrees of freedom of the entire system, rendering it inaccessible to local measurements.
Scrambling is also related to the Heisenberg dynamics of local operators, and can be probed via the squared commutator of two local and Hermitian operators $W_1,V_r$, at positions $1$ and $r$ respectively,
\begin{equation}
\label{eq:square-comm}
\mathcal{C}(r,t) =-\frac{1}{2}\expval*{\comm*{W_1(t)}{V_r}^2},
\end{equation}
where $W_1(t)$ is the Heisenberg evolved operator. The growth of the squared commutator corresponds to $W_1(t)$ increasing in size and complexity, leading it to fail to commute with $V_r$.
In a local quantum chaotic system, $\mathcal{C}(r,t)$ typically spreads ballistically, exhibiting rapid growth ahead of the wavefront and saturation behind, at late times \cite{Patel2017,Xu2019,Xu2020}.

Of particular interest are the so-called fast scramblers, systems where $\mathcal{C}(r,t)$ reaches $O(1)$ for all $r$ in a time $t_s \propto \log(N)$, with $N$ being the number of degrees of freedom.
Among the best known examples are black holes, which are conjectured to be the fastest scramblers in nature \cite{Sekino2008,Lashkari2013,Shenker2014,Maldacena2016}, as well as the Sachdev-Ye-Kitaev (SYK) \cite{Sachdev1993,Kitaev2015} model and other related holographic models \cite{Danshita2017,Chen2018c,Chew2017,Gu2017a}.

Recent advances in the development of coherent quantum simulators have enabled the study of out-of-equilibrium dynamics of spin models with controllable interactions \cite{Cirac2012}, making them ideal platforms to experimentally study information scrambling. 
Several experiments have already been performed \cite{Li2017,Garttner2017,Wei2018a,Landsman2019,Joshi2020,Blok2020}, probing scrambling in either local or non-chaotic systems.
The experimental observation of fast scrambling remains challenging however, particularly because few systems are known to be fast scramblers, and those that are, like the SYK model, are highly non-trivial, involving random couplings and many-body interactions.
Some recent proposals suggested that spin models with non-local interactions can exhibit fast scrambling \cite{Swingle2016,Marino2019,Bentsen2019}, albeit with complicated and inhomogeneous interactions.

In this paper, we argue that the simplest possible global interaction, together with chaotic dynamics, are sufficient to make a spin model fast scrambling.
We consider spin-1/2 chains with Hamiltonians of the form
\begin{equation}
\label{eq:Ham-generic}
\mathcal{H} = \mathcal{H}_{\text{local}} - \frac{g}{\sqrt{N}}\sum_{i<j}Z_iZ_j,
\end{equation}
where $Z_i$ is the Pauli $z$ operator acting on site $i$ and 
$\mathcal{H}_{\text{local}}$ is a Hamiltonian with only local interactions that ensures that the full $\mathcal{H}$ is chaotic.
We note that such global interactions are ubiquitous in ultracold atoms in optical cavities \cite{Sorensen2002,Chaudhury2007,Fernholz2008,Leroux2010,Schleier-Smith2010}, and also in ion traps \cite{Sawyer2012,Britton2012,Wang2013,Bohnet2016}.

We first show that this effect is generic, by studying two models, a random quantum circuit and a classical model, both designed to mimic the universal dynamics of \cref{eq:Ham-generic}.
We then provide numerical evidence for fast scrambling for a particular time-independent quantum Hamiltonian.
Finally, we discuss possible experimental realizations.

\begin{figure}
	\centering
	\includegraphics[width=\linewidth]{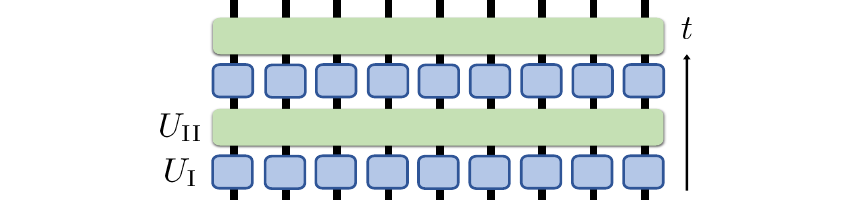}
	\caption{Diagram of the random circuit. As given in \cref{eq:two-layer-circuit-def}, each blue square is an independent Haar-random unitary $U_{H,i}$ acting on site $i$, and the green rectangle is the global interaction $\Uzz$.}
	\label{fig:circuit}
\end{figure}

{\it Random circuit model.---}%
As a proof-of-principle, we consider a system of $N$ spin-1/2 sites, with dynamics generated by a random quantum circuit (see \cref{fig:circuit}) inspired by the Hamiltonian in \cref{eq:Ham-generic}.
While less physical than the Hamiltonian model, it has the advantage of being exactly solvable while providing intuition about generic many-body chaotic systems with similar features. 

The time-evolution operator is $U(t) = (\Uzz\Uhar)^t$ where a single-time-step update consists of the two layers
\begin{equation}
\label{eq:two-layer-circuit-def}
\Uhar = \prod_{i=1}^N U_{H,i},\qquad \Uzz = e^{-i\frac{g}{2\sqrt{N}}\sum_{i<j}Z_iZ_j},
\end{equation}
where each $U_{H,i}$ is an independent Haar-random single-site unitary.
The two layers in \cref{eq:two-layer-circuit-def} are motivated by the two terms in \cref{eq:Ham-generic}, with the Haar-random unitaries replacing $\mathcal{H}_{\text{local}}$.

We are interested in the operator growth of an initially simple operator $\Oc$.
At any point in time, the Heisenberg operator $\Oc(t) =U^\dagger(t)\Oc U(t)$ can be decomposed as $\Oc(t) = \sum_\Sc a_{\Sc}(t)\Sc$, where $\Sc$ is a string composed of the Pauli matrices and the identity, forming a basis for $SU(2^N)$.
As in random brickwork models \cite{Nahum2018,VonKeyserlingk2018} and random Brownian models \cite{Xu2019}, the Haar-averaged probabilities $\expval{a_{\Sc}^2(t)}$, encoding the time evolution of $\Oc(t)$, themselves obey a linear equation
\begin{equation}
\label{eq:a_S_2-haar-general}
\expval{a_{\Sc}^2(t+1)} = \sum_{\Sc'}W_{\Sc,\Sc'}\expval{a_{\Sc'}^2(t)}.
\end{equation}
Here, $W_{\Sc,\Sc'}$ is a $4^N\times 4^N$ stochastic matrix describing a fictitious Markov process \cite{Dahlsten2007,Znidaric2008}. 
The average probabilities $\expval{a_{\Sc}^2(t)}$ fully determine the growth of the average of $\Cc(t)$ in \cref{eq:square-comm} (see Supplemental Material (SM) \cite{supp}). 
Because of the Haar unitaries and the simple uniform interaction in \cref{eq:two-layer-circuit-def}, $W_{\Sc,\Sc'}$ is highly degenerate and only depends on the total weights of the strings $\Sc,\Sc'$, counting the number of non-identity operators, i.e $w(\Sc) = \sum_i (1-\delta_{\Sc_i,1})$, and on the number of sites where both $\Sc$ and $\Sc'$ are non-identity, i.e $v(\Sc,\Sc') = \sum_i(1-\delta_{\Sc_i,1})(1-\delta_{\Sc_i',1})$, and is given by (see SM for derivation \cite{supp}) \footnote{The term with $2l=k,w+w'-2v=0$ is assumed to be $0^0=1$. See \cite{supp}.}
\begin{align}
\label{eq:W-def}
&W(w,w',v) =  
\qty(\frac{1}{3})^{w+w'}\sum_{k=0}^{v}\mqty(v\\k)\sum_{l=0}^{k}\mqty(k\\l)\times\\ \nonumber
&\qty[\cos[2](\frac{2l-k}{\sqrt{N}}g)]^{N-k-(w+w'-2v)}\qty[\sin[2](\frac{2l-k}{\sqrt{N}}g)]^{w+w'-2v}.
\end{align}
If we further assume that $\Oc$ starts out as a single site operator on site $1$, then throughout the evolution, $\expval{a_{\Sc}^2(t)}$ only depend on the total operator weight $w$, and the weight on site $1$, which we denote by $w_1\in\{0,1\}$. We thus introduce the operator weight probability $h_t$ at time $t$, 
\begin{equation}
h_t(w,w_1) = \expval{a_{\Sc}^2(t)} 3^w\mqty(N-1\\w-w_1),
\end{equation} 
which gives the probability of $\Oc(t)$ having total weight $w$ and weight $w_1$ on site 1.

\begin{figure}
	\centering
	\includegraphics[width=\linewidth]{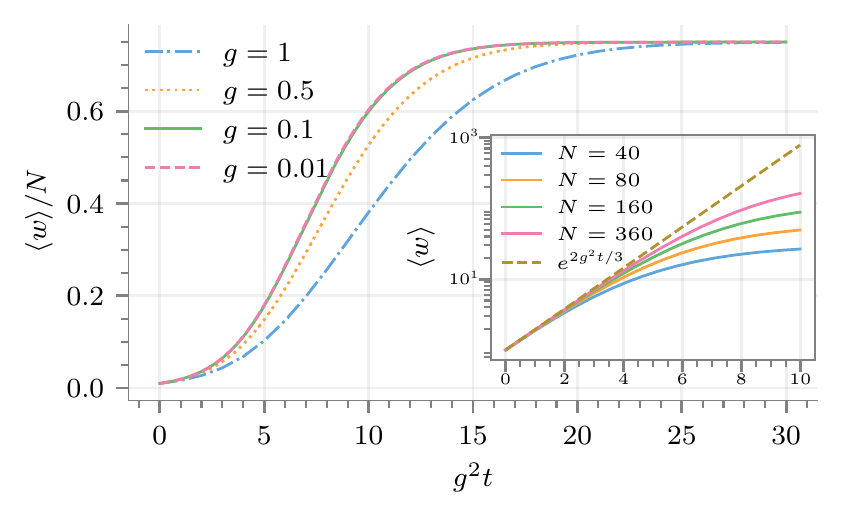}
	\caption{Normalized mean operator weight $\expval{w(t)}/N = \frac{1}{N}\sum_w w\,h_t(w)$ as a function of time for different $g$ and $N=100$, computed using \cref{eq:htilde-master-eq}. For small enough $g$, all the curves collapse to a single curve as a function of $g^2t$, as implied by \cref{eq:Fokker-Planck}. The inset shows the initial exponential increase of $\expval{w(t)}$ for different system sizes $N$ and $g=0.1$.}
	\label{fig:meanw}
\end{figure}

The time evolution of $h_t(w,w_1)$ is given by the master equation
\begin{equation}
\label{eq:htilde-master-eq}
h_{t+1}(w,w_1) = \sum_{w_1'=0,1}\sum_{w'=w_1'}^{N-1+w_1'}\mathcal{R}(w,w_1,w',w_1')h_{t}(w',w_1'),
\end{equation}
where the $2N\times 2N$ matrix $\mathcal{R}$ is 
\begin{align}
\label{eq:tildeW-def}
\mathcal{R}(w,w_1,&w',w_1') = 
3^{w}\sum_{m=0}^{\min\{w-w_1,w'-w_1'\}}\mqty(w'-w_1'\\m)\\
&\mqty(N-1-w'+w_1'\\w-w_1-m) W(w,w',m+w_1w_1'). \nonumber
\end{align}
The transition matrix $\mathcal{R}$, scaling only linearly with $N$, allows us to efficiently simulate the dynamics for large system sizes (see \cref{fig:meanw}).

To proceed analytically, we Taylor-expand \cref{eq:W-def} to leading order in $g$, which gives rise to a closed master equation for the total operator weight probability $h_t(w) \equiv h_t(w,0)+h_t(w,1)$,
\begin{align}
\label{eq:h-approx-master-eq}
&\frac{h_{t+1}(w)-h_t(w)}{g^2} = \frac{2w}{9N}(1-3N+2w)h_t(w)\\
&+\frac{2w(w+1)}{9N}h_t(w+1) +\frac{N-w+1}{3N}2(w-1)h_t(w-1),\nonumber
\end{align}
which is similar to random Brownian models \cite{Xu2019,Zhou2019} and shows that, at $O(g^2)$, $w$ can change by at most $\pm1$ in a single step.
Assuming that $h(w,t)$ varies slowly with respect to $g^2t$ and $w$, we can approximate the above equation by a Fokker-Planck equation (rescaling time $\tau = g^2 t$)
\begin{equation}
\label{eq:Fokker-Planck}
\partial_\tau h(w,\tau) = -\partial_w\qty(D_1(w)h(w,\tau))
+\partial_w^2\qty(D_2(w)h(w,\tau)),
\end{equation}
where the drift and diffusion coefficients are (dropping higher order terms $O(1/N,w/N)$)
\begin{equation}
D_1(w) =  \frac{2}{3}\qty(w-\frac{4w^2}{3N}),\qquad D_2(w) = \frac{w}{3}-\frac{2w^2}{9N}.
\end{equation}

This equation describes the rapid growth of an initially localized distribution, followed by a broadening and finally saturation (see \cref{fig:meanw} and SM \cite{supp} for more details). At early time, the $\frac{2}{3}w$ term in the drift coefficient dominates, giving rise to exponential growth of the mean operator weight $\expval{w(t)}\sim e^{2g^2t/3}$, which agrees with the full numerical solution of the master equation, as can be seen in \cref{fig:meanw}. The mean weight is related to the infinite-temperature squared-commutator in \cref{eq:square-comm} (averaged over different circuits) via $\expval{\mathcal{C}(t)} = \frac{4}{3}\expval{w(t)}/N$ \cite{supp}.
Since $\expval{w(t)}$ grows exponentially with time, $\expval{w(t)}$ reaches $O(N)$ and $\expval{\mathcal{C}(t)}$ reaches $O(1)$ when $t=\frac{3}{2g^2} \log(N)$, thus establishing that this model is fast scrambling. 
Note that the $1/\sqrt{N}$ normalization in \cref{eq:Ham-generic,eq:two-layer-circuit-def} is crucial.
Had we chosen instead $1/N\, (g\rightarrow g/\sqrt{N})$, the Lyapunov exponent would have been $\frac{2g^2}{3N}$ and the scrambling time would have been $t\sim N\log(N)$.

{\it Classical Model.---}%
Let us now consider a different setting that also allows to probe the basic timescales involved, and shows that randomness is not required. A convenient tractable choice is a classical model consisting of globally coupled non-linear oscillators. Note that the analogs of out-of-time-order correlators (OTOCs) have been studied in a variety of classical models~\cite{Rozenbaum_2017,PhysRevLett.121.250602,PhysRevLett.122.024101,Jalabert_2018,Yan2020a,Marino2019}
and have been shown to capture the scrambling dynamics of quantum models like the SYK model  \cite{Kurchan2018,Schmitt2019,Scaffidi2019}. 

Consider a $2N$-dimensional phase space with coordinates $q_r$ (positions) and $p_r$ (momenta) for $r=1,\cdots,N$ with canonical structure specified by the Poisson brackets $\{ q_r , p_s \}_{PB} = \delta_{rs}$.
The Hamiltonian is $\mathcal{H}_c = K + V_2 + V_4$ where
\begin{align}
K &= \sum_{r=1}^N \frac{p_r^2}{2},\qquad V_4= \frac{\Omega_3^2}{4} \sum_{r=1}^N q_r^4,\\
V_2 &=  \frac{\Omega_1^2}{2} \sum_{r=1}^{N-1} (q_{r+1}-q_r)^2 + \frac{\Omega_2^2 }{2\sqrt{N}} \left( \sum_{r=1}^N q_r \right)^2.
\end{align}

\begin{figure}[htb]
	\centering
	\includegraphics[width=\linewidth]{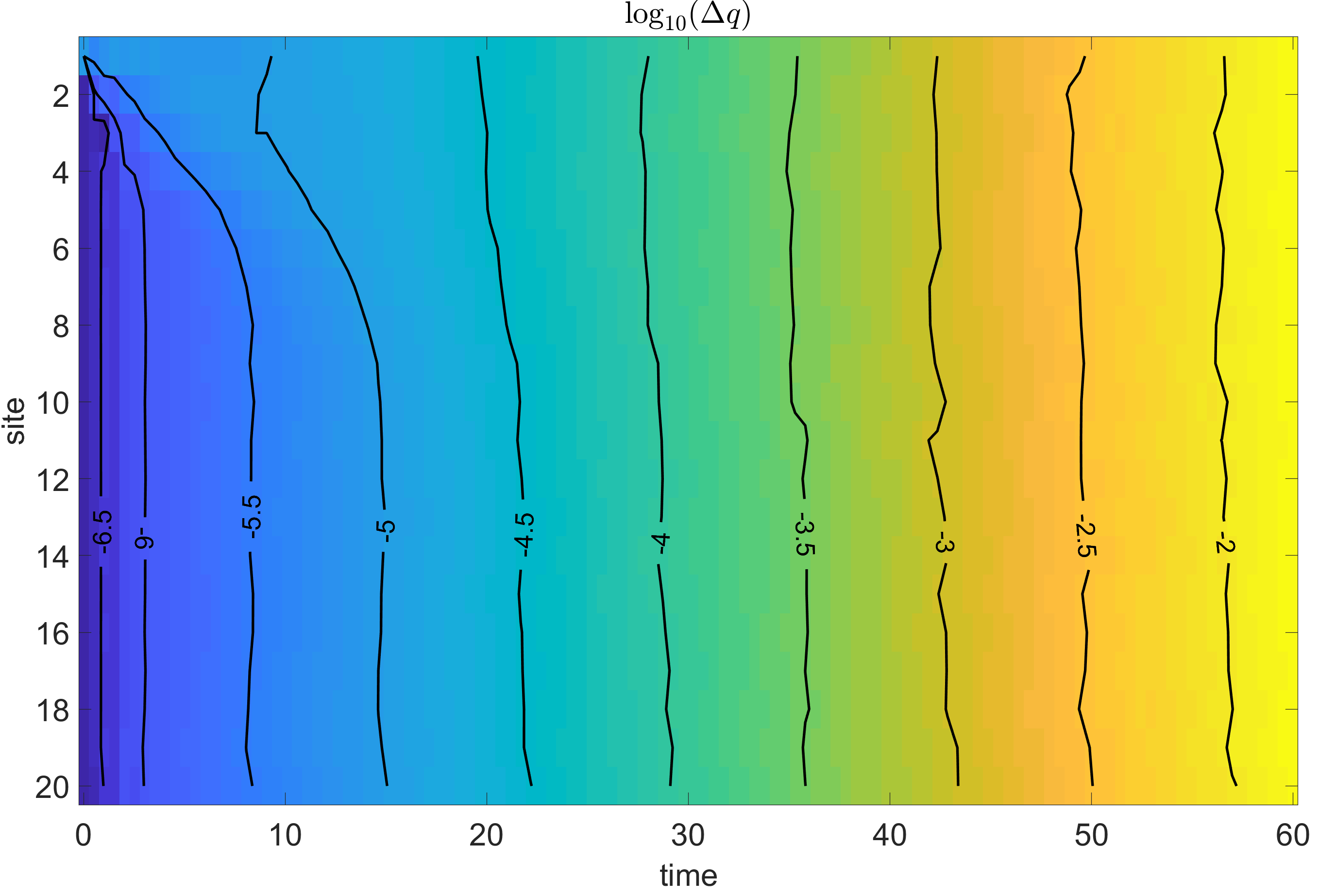}
	\caption{$\log_{10} \Delta q_r(t)$ for $N=20$, $\epsilon=10^{-5}$, $\Omega_1=\Omega_2=1$, and $\Omega_3=2$. The labeled black lines are contours of constant $\log_{10} \Delta q$. Early time ballistic growth is visible in the upper left corner while at later times the system exhibits spatially uniform exponential growth in time.
	}
	\label{fig:classical}
\end{figure}

The timescales for the growth of perturbations under $\mathcal{H}_c$ dynamics may be understood in two stages. First, $K+V_2$ can be solved exactly; this combination of terms provides the non-locality. The remaining $V_4$ term renders the dynamics chaotic, provided $\Omega_3$ is large enough. The dynamics of $K+V_2$ causes a localized perturbation to spread to every oscillator with non-local amplitude $1/N$ in a time of order $1/N^{1/4}\Omega_2$. Then conventional local chaos can amplify this $1/N$-sized perturbation to order-one size in a time of order $\lambda ^{-1} \ln N$, where $\lambda $ is some typical Lyapunov exponent.

At the quadratic level, the uniform mode, $Q = \frac{1}{N} \sum_r q_r$,
is decoupled from the remaining modes of the chain. 
Hence, the propagation of any perturbation is a superposition of the motion due to the local $\Omega_1$ terms and the special dynamics of the uniform mode. Since the local terms cannot induce non-local perturbations, we may focus on the dynamics of the uniform mode.
The uniform mode's equation of motion is $\frac{d^2 Q}{dt^2} = - \sqrt{N}\Omega_2^2 Q$
with solution
\begin{equation}
Q(t) = Q(0) \cos(N^{\frac{1}{4}}\Omega_2 t) + \frac{\frac{dQ}{dt}(0)}{N^{\frac{1}{4}}\Omega_2} \sin(N^{\frac{1}{4}}\Omega_2 t).
\end{equation}
A localized perturbation on site $1$ with zero initial time derivative can be written as $\delta \va{q}(0) = \epsilon \left(  [\vu{e}_1 - \vu{u}_0] + \vu{u}_0 \right)$,
where $\vu{u}_0 = [1, \cdots, 1]^T/N$ represents the uniform mode, $\vu{e}_1 = [1, 0, \cdots, 0]^T$, and $\vu{e}_1 - \vu{u}_0$ is orthogonal to the uniform mode. 
The orthogonal mode evolves in a local fashion, hence $\delta \va{q}(t) = \epsilon \left( \text{local piece}  + \vu{u}_0 \cos(N^{\frac{1}{4}}\Omega_2 t) \right)$.
For oscillators far from the initial local perturbation, the dynamics is given by 
\begin{equation}
\delta q_{r\gg 1}(t) = \frac{\epsilon}{N} \qty[\cos(N^{\frac{1}{4}}\Omega_2 t) - 1].
\end{equation}
Thus, after a time $\pi/N^{\frac{1}{4}}\Omega_2$, any localized perturbation has spread to distant sites with amplitude $\epsilon/N$.

The inclusion of $V_4$ renders the equations of motion non-linear and the system chaotic in at least part of the phase space. We leave a detailed study of the classical chaotic dynamics of this model to the future, but as can be seen in \cref{fig:classical}, a numerical solution of the equations of motion displays sensitivity to initial conditions.

The precise protocol is as follows. We compare the dynamics of two configurations, $\va{q}^{(1)}$ and $\va{q}^{(2)}$, averaged over many initial conditions. The initial condition of configuration one has each oscillator start at rest from a random amplitude drawn uniformly and independently from $[-1,1]$. Configuration two is identical to configuration one except that $q^{(2)}_1(0) = q^{(1)}_1(0)+\epsilon$ for $\epsilon =10^{-5}$. Both configurations are evolved in time and the difference $\Delta q_r(t) = |q^{(2)}_r(t)-q^{(1)}_r(t)|$ is computed and averaged over $4000$ different initial conditions.
\Cref{fig:classical} shows this average of $\Delta q_r$ for $N=20$ with $\Omega_1=1$, $\Omega_2=1$, and $\Omega_3=2$. Because the system can generate an $\epsilon/N$-sized perturbation on all sites in a short time, the subsequent uniform exponential growth implies that any local perturbation will become order one on all sites after a time $\sim \lambda^{-1} \log \frac{N}{\epsilon}$.

The above analysis corresponds to the classical limit of coupled quantum oscillators where some effective dimensionless Planck's constant vanishes, $\hbar_{\text{eff}} \rightarrow 0$. In the opposite limit of large $N$ at fixed $\hbar_{\text{eff}}$, the dynamics of quantum OTOCs can be obtained from the corresponding classical Lyapunov growth up to a timescale of order $\log \frac{1}{\hbar_{\text{eff}}} \ll \log N$. At later times, one needs to consider fully quantum local dynamics. If one imagines breaking the system up into local clusters and if each cluster can be viewed as a quantum chaotic system with random-matrix-like energy levels, a dynamical system not unlike the random circuit model above is obtained.

{\it Chaos and level statistics.---}%
Having established fast scrambling in both the random circuit and the classical model, we now return to the quantum spin model of \cref{eq:Ham-generic}.
We first examine whether such a model is chaotic, which is a necessary condition for it being fast scrambling.
For the local Hamiltonian part, we consider the mixed-field Ising chain 
\begin{equation}
\label{eq:H-local}
\mathcal{H}_{\text{local}} = -J\sum_iZ_iZ_{i+1} -h_x\sum_iX_i-h_z\sum_iZ_i.
\end{equation} 
A standard approach to identify a transition from integrability to quantum chaos is based on a comparison of energy-level-spacing statistics with Poisson and Wigner-Dyson distributions. Another convenient metric is the average ratio of consecutive level spacings \cite{Bogomolny2013} $\langle r\rangle$, where $r =  \min{(r_n, 1/r_n)}$, $r_n=\delta_n/\delta_{n-1}$, $\delta_n = E_n-E_{n-1}$, and $E_n$ are the eigenvalues ordered such that $E_n \geq E_{n-1}$.

As was already suggested in Ref.\ \cite{Lerose2018} for a similar model, we find that the longitudinal field is unnecessary, and the full system can have Wigner-Dyson statistics even for $h_z=0$, in which case $\mathcal{H}_{\text{local}}$ is integrable. 
The resulting Hamiltonian reads 
\begin{equation}\label{eq:H_ZZ_tot}
\mathcal H = -J\sum_iZ_i Z_{i+1} -h_x\sum_iX_i-\frac{g}{\sqrt{N}}\sum_{i<j} Z_i Z_j.
\end{equation}

\begin{figure}
	\centering
	\includegraphics[width=\linewidth]{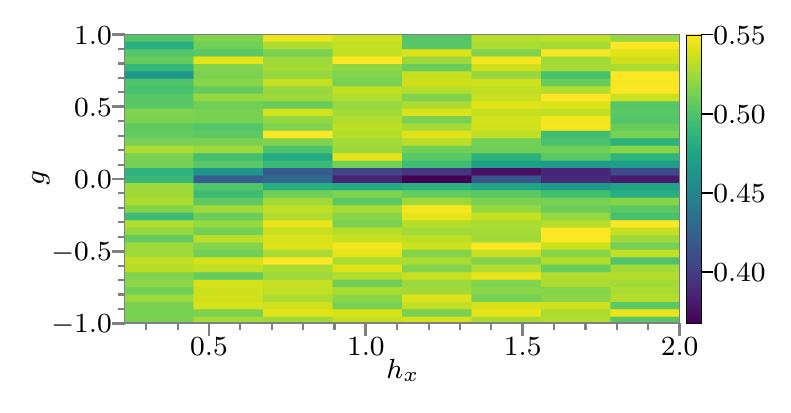}
	\caption{Average adjacent-level-spacing ratio $\langle r \rangle$ for the model in \cref{eq:H_ZZ_tot} with $J=1$. 
		Data corresponds to a system of $N=15$ spins with periodic boundary conditions for fixed momentum and $Z$-reflection symmetry blocks of the Hamiltonian.}
	\label{fig:r_av}
\end{figure}

Average adjacent-level-spacing ratio changes from $\langle r\rangle_{Pois}\approx 0.38$ for Poisson level statistics to $\langle r\rangle_{GOE}\approx 0.53$ for Wigner-Dyson level statistics in the Gaussian Orthogonal Ensemble (GOE) \cite{Bogomolny2013}. 
In the vicinity of $g\to 0$, $\expval{r}$ (see Fig.\ \ref{fig:r_av}) shows proximity to Poisson statistics, while, for $|g|\gtrsim 0.25$, the level statistics agree with those of the GOE.

{\it Out-of-time-order correlator and entanglement growth.---}%
We now study the dynamics of an OTOC and entanglement entropy in the spin chain.
We consider the following OTOC
\begin{equation}
\label{eq:OTOC}
F(r,t)  =\Re[\expval{Z_1(t)Z_rZ_1(t)Z_r}],
\end{equation}
which is related to \cref{eq:square-comm} by $\Cc(r,t)=1-F(r,t)$.
The expectation value is evaluated in a Haar-random pure state,  which approximates the infinite-temperature OTOC, but enables us to reach larger system sizes \cite{Luitz2017}.

\begin{figure}
	\centering
	\includegraphics[width=\linewidth]{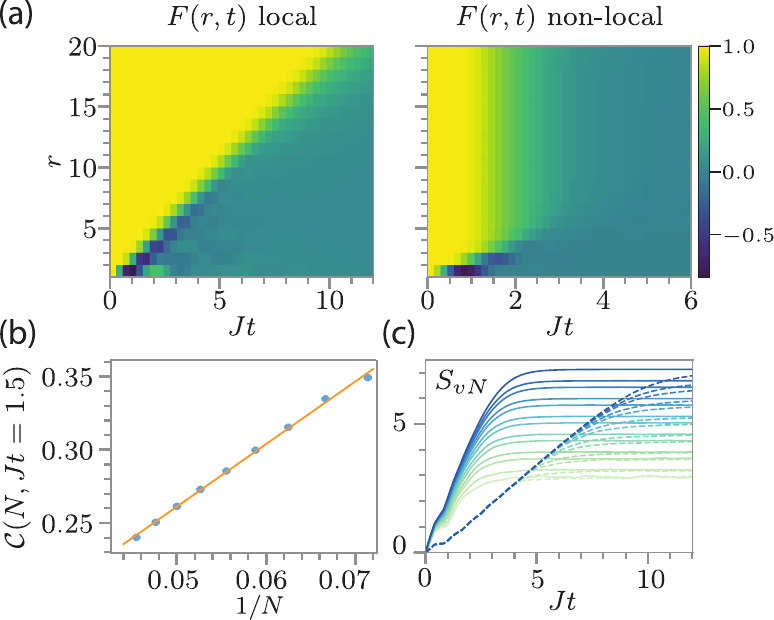}
	\caption{(a) Time evolution of the OTOC for the (left) local and (right) non-local models. (b) $1-F(N,t)$ after a fixed evolution time for the non-local model (for different system sizes $N$), showing a linear dependence on $1/N$. The orange line is a linear fit. (c) Half-cut entanglement-entropy growth starting from the $+\vu{y}$ state for local (dashed) and non-local (solid) models. The color indicates the system size, starting from $N=10$ (light green) until $N=22$ (dark blue). For all plots,  $J=1,h_x=1.05$ and $h_z=0.5,g=0$ ($h_z=0,g=-1$) for the local (non-local) models.}
	\label{fig:otoc_fig}
\end{figure}

In \cref{fig:otoc_fig}(a), we show the OTOC for an open chain of $N=20$ spins for both the local model, governed by $\mathcal{H}_{\text{local}}$ only, and the non-local model in \cref{eq:H_ZZ_tot}, which includes the global interaction.
In the local case, the OTOC spreads ballistically, forming a linear light cone. 
In contrast, in the non-local case, the spreading is super-ballistic and $F(r\gg 1,t)$ is approximately independent of $r$, as expected for a fast scrambler.
As we discussed in the context of the classical model, a necessary condition for fast scrambling is that, before the onset of exponential growth, the decay of correlations with $N$ should be at most algebraic ($\mathcal{C}\propto N^{-\alpha}$) and not exponential. 
In \cref{fig:otoc_fig}(b), we verify that this is the case for the non-local model, showing that $\mathcal{C}\propto N^{-1}$ between the two ends of the chain after a fixed time.

\Cref{fig:otoc_fig}(c) shows the half-cut entanglement entropy following a quench starting from the $+\vu{y}$ state for both models. For the local model, the entanglement grows linearly in time before saturating,  whereas the non-local model shows a significant speed up.
Moreover, in the non-local model, the growth rate clearly increases with the system size, further supporting our claim.

{\it Experimental realization.---}%
The Hamiltonian in \cref{eq:H_ZZ_tot}, and many variations of it, can be experimentally realized in a variety of platforms.
A natural realization is with Rydberg dressing of neutral atoms~\cite{Honer2010,Grass2018,Pupillo2010,Glaetzle2012}.
The spin can be encoded in two ground states with one of them dressed to two Rydberg states such that one of the Rydberg states leads to all-to-all interactions and the second to nearest-neighbor interactions.
Other similar spin models can be realized with cavity-QED setups, using photon-mediated all-to-all interactions~\cite{Sorensen2002,Leroux2010,Gopalakrishnan2011,Strack2011} of the  
$XX$ or $XXZ$-Heisenberg form~\cite{Swingle2016,Bentsen2019} together with nearest-neighbor interactions achieved by Rydberg-dressing one of the grounds states~\cite{Gelhausen2016,Zhu2019a}. 
Other possibilities include a chain of coupled superconducting qubits, with all-to-all flip-flop interactions mediated via a common bus~\cite{Majer2007,Zhu2016c,Onodera2019} or trapped ions \cite{Sawyer2012,Britton2012,Wang2013,Genway2014,Bohnet2016}.

{\it Conclusion and outlook.---}%
In this paper, we argued that a single global interaction together with local chaotic dynamics is sufficient to give rise to fast scrambling.
While fast scrambling is intrinsically difficult to study numerically, our numerical evidence, together with the semi-classical analysis and the exactly solvable random circuit, provide a compelling argument in favor of our claim.
Our models do not require disordered or inhomogeneous couplings and are within reach of current state-of-the-art quantum simulators. 
Thus, an experimental implementation of the spin model could test our claims on much larger systems sizes, something that may very well be impossible to do on a classical computer. This can pave the way towards experimental investigations of aspects of quantum gravity.

Future theoretical work may include a more systematic analysis of the $N$-dependence of various timescales, e.g. for entanglement growth, and of the behaviour of the OTOC at low temperatures. It is also interesting to investigate whether similar conclusions can be reached without perfectly uniform global interactions, for example with power-law decaying interactions.

\begin{acknowledgments}
	We thank A.\ Lucas for useful discussions, especially regarding the normalization of the global interactions.
	We acknowledge discussions with S. Xu at early stages of this work and P. Titum for pointing out the Hamiltonian [\cref{eq:H_ZZ_tot}] to us.
	R.B., P.B., Y.A.K., and A.V.G.\ acknowledge funding by the NSF PFCQC program, DoE ASCR Quantum Testbed Pathfinder program (award No.~DE-SC0019040), DoE BES Materials and Chemical Sciences Research for Quantum Information Science program (award No.~DE-SC0019449), DoE ASCR Accelerated Research in Quantum Computing program (award No.~DE-SC0020312), AFOSR MURI, AFOSR, ARO MURI, NSF PFC at JQI, and ARL CDQI. R.B.\ acknowledges support of NSERC and FRQNT of Canada. B.G.S. is supported in part by the U.S. Department of Energy, Office of Science, Office of High Energy Physics QuantISED Award de-sc0019380.\\
	\textit{Note added.} We would like to draw the reader's attention to two related parallel works which appeared recently: by  Li,  Choudhury,  and Liu \cite{Li2020}, on fast scrambling with similar spin models; and by Yin and Lucas \cite{yin2020bound}, on lower bounds of the scrambling time in similar spin models.
\end{acknowledgments}


%

\end{document}


\title{Supplemental Material: Minimal Model for Fast Scrambling}
\date{\today}
\author{Ron Belyansky}
\author{Przemyslaw Bienias}
\author{Yaroslav A. Kharkov}
\author{Alexey V. Gorshkov}
\author{Brian Swingle}
\affiliation{Joint Center for Quantum Information and Computer Science, NIST/University of Maryland, College Park, Maryland 20742 USA}
\affiliation{Joint Quantum Institute, NIST/University of Maryland, College Park, Maryland 20742 USA}

\maketitle
\onecolumngrid

\renewcommand{\theequation}{S\arabic{equation}}
\renewcommand{\thesubsection}{S\arabic{subsection}}
\renewcommand{\thesubsubsection}{\Alph{subsubsection}}

\renewcommand{\bibnumfmt}[1]{[S#1]}
\renewcommand{\citenumfont}[1]{S#1} 

\pagenumbering{arabic}

\makeatletter
\renewcommand{\thefigure}{S\@arabic\c@figure}
\renewcommand \thetable{S\@arabic\c@table}

In this Supplemental Material we present additional details concerning the random circuit. 
In \cref{sec:W-matrix}, we derive the general transition rate matrix $W$, given in Eq.~(5) of the main text. In \cref{sec:master-eq} we specialize it to the case of an initial single-site operator, deriving  Eqs.~(7) and (8) of the main text. 
In \cref{sec:cont-approx}, we present the continuum approximation for small $g$, deriving the Fokker-Planck equation, Eqs.~(9-11) of the main text.
In \cref{sec:squared-comm-w}, we derive the relation between the average squared commutator and the mean operator weight.
In \cref{sec:dynamics}, we provide additional details on the dynamics and steady-state of the probability weight distribution.
In \cref{sec:onestep}, we derive an analytical expression for the probability weight distribution after one step of the random circuit and show that if the interactions are strong enough, the scrambling time is $O(1)$.

\section{Derivation of the stochastic matrix $W$}
\label{sec:W-matrix}

To be slightly more general, we consider a system of $N$ sites, each of local dimension $q$.
As discussed in the main text, we are interested in the time evolution of a simple initial operator $\Oc(t) =U^\dagger(t)\Oc U(t)$ 
\begin{equation}
\label{eq:O-heisenberg-expansion-strings}
\Oc(t) = \sum_\Sc a_{\Sc}(t)\Sc,
\end{equation}
where the strings $\Sc$ form a basis for $SU(q^N)$, normalized as $\tr(\Sc)=q^N\delta_{\Sc,1},\tr(\Sc \Sc')=q^N\delta_{\Sc\Sc'}$.
We take $U(t) = \prod_{i=1}^tU_i$ where $U_i = \Uhar\Uzz\Uhar$ and
$\Uhar$ is a product of single site Haar random unitaries while $\Uzz$ is the global interaction. Note that the two $\Uhar$ appearing on either side of the $\Uzz$ are different, i.e the random unitaries are random in both space in time.
Here we inserted an additional layer of the Haar unitaries, as compared to the main text.
This is completely equivalent, as this extra layer can always be absorbed into the Haar layer of either the step before or the step after, but it simplifies calculations.

Using $a_\Sc(t)=q^{-N}\tr(\Oc(t)\Sc)$, we can write $a_\Sc^2(t)$ in terms of the coefficients at the previous time step
\begin{dmath}
	\label{eq:a2-general-eq}
	a^2_\Sc(t) = q^{-2N}\sum_{\Sc',\Sc''}a_{\Sc'}(t-1)a_{\Sc''}(t-1)\tr(U^\dagger\Sc'U\Sc)\tr(U^\dagger\Sc''U\Sc).
\end{dmath}

Thus, we want to evaluate the quantity 
\begin{dmath}
	\expval{\tr(U^\dagger\Sc'U\Sc)\tr(U^\dagger\Sc''U\Sc)},
\end{dmath}
where $\expval{...}$ denotes Haar average over the random unitaries. 

Using properties of trace, we can write
\begin{dmath}
	\label{eq:trace-first}
	\expval{\tr(U^\dagger\Sc'U\Sc)\tr(U^\dagger\Sc''U\Sc)}=\expval{\tr(U^\dagger\Sc'U\Sc\otimes U^\dagger\Sc''U\Sc)}.
\end{dmath}
In doing so, we now have a trace over two copies of the system, which could still be thought as a $N$-site system, where every site is now of dimension $q^2$ instead of $q$.
In the following, we will denote operators acting on the right system by an overbar. For example $Z_i\bar{Z_i}$ corresponds to the Pauli $Z$ operator acting on site $i$ of both copies, i.e $Z_i\otimes Z_i$.

For our choice of $U$, \cref{eq:trace-first} becomes
\begin{dmath}
	\label{eq:two-traces-general-one-time-step}	\expval{\tr(U^\dagger\Sc'U\Sc)\tr(U^\dagger\Sc''U\Sc)}
	=
	\tr(\expval{(\Uhar\otimes \Uhar)(\Uzz\otimes \Uzz)\expval{\Uhar\Sc' \Uhar^\dagger\otimes \Uhar\Sc'' \Uhar^\dagger}(\Uzz\otimes \Uzz)^\dagger(\Uhar\otimes \Uhar)^\dagger}(\Sc\otimes\Sc)).
\end{dmath}

We will calculate the above in several steps, working from inside out
\begin{align}
\Ic_1 &= \expval{\Uhar\Sc' \Uhar^\dagger\otimes \Uhar\Sc'' \Uhar^\dagger},\\
\Ic_2 &= (\Uzz\otimes \Uzz)\Ic_1(\Uzz\otimes \Uzz)^\dagger,\\
\Ic_3 &= \expval{(\Uhar\otimes \Uhar)\Ic_2(\Uhar\otimes \Uhar)^\dagger},
\end{align}
with $\tr(\Ic_3(\Sc\otimes\Sc))$ being our quantity of interest.

Before proceeding, let us introduce an important formula for calculating the Haar averages. 
Consider a $d^2\times d^2$ matrix $A$, and a $d\times d$ Haar random unitary matrix $U$. Then, we have the following formula \cite{Zhang2014,Nahum2018}
\begin{dmath}
	\label{eq:Haar-average-formula}
	\expval{(U\otimes U)A(U\otimes U)^\dagger} \equiv \int_{U(d)}(U\otimes U)A(U\otimes U)^\dagger d\mu(U)=
	\qty(\frac{\tr(A)}{d^2-1}-\frac{\tr(AF)}{d(d^2-1)})\idop_{d^2}-\qty(\frac{\tr(A)}{d(d^2-1)}-\frac{\tr(AF)}{d^2-1})F,
\end{dmath} 
where $F = \sum_{ij}\ket{ij}\bra{ji}$ is the swap operator.

From this, it follows that 
\begin{equation}
\label{eq:Haar-average-general}
\Ic_1 = 
\prod_r\expval{U_r^\dagger\Sc_r'U_r\otimes U_r^\dagger\Sc_r''U_r}
=
\delta_{\Sc',\Sc''}\prod_r\qty(\frac{q^{2}\delta_{\Sc_r',1}-1}{q^{2}-1}\idop_{q^{2}}+\frac{q-q\delta_{\Sc_r',1}}{q^{2}-1}F_r)
\end{equation}
where we used $\tr(\Sc_r)=q\delta_{\Sc_r,1}$ and $\tr(\Sc_r\Sc_r')=q\delta_{\Sc_r,\Sc_r'}$.
Here $F_r$ swaps site $r$ of the left system with the corresponding site $r$ of the right system.

The overall delta function $\delta_{\Sc',\Sc''}$ immediately implies that the Haar average of \cref{eq:a2-general-eq} may be written as 
\begin{dmath}
	\label{eq:a_S_2-haar-general}
	\expval{a_{\Sc}^2(t+1)} = \sum_{\Sc'}W_{\Sc,\Sc'}\expval{a_{\Sc'}^2(t)},
\end{dmath}
where $W_{\Sc,\Sc'} = q^{-2N}\tr(\Ic_3(\Sc\otimes\Sc))$.

To proceed, we specialize to qubits, i.e. $q=2$, in which case the swap operator can be written as $F_r=\frac{1}{2}(\idop_r\otimes \bar{\idop}_r + \bm{\sigma}_r\cdot\bar{\bm{\sigma}}_r) = \frac{1}{2}(\idop_r \bar{\idop}_r+X_r\bar{X}_r+Y_r\bar{Y}_r+Z_r\bar{Z}_r)$ where bar denotes operators acting on the second system. We can combine all the $\idop$s together, giving
\begin{dmath}
	\Ic_1=\label{eq:Haar-qubits-general}
	\delta_{\Sc',\Sc''}\prod_i\qty(\delta_{\Sc_i',1}\idop_{2^{2}}+\frac{1-\delta_{\Sc_i',1}}{3}\bm{\sigma}_i\cdot\bar{\bsigma}_i) = 	\delta_{\Sc',\Sc''}\sum_{\Omega_{\Sc'} \subset\{1,2,\cdots,N\}}\prod_{i\in\{1,2,\cdots,N\}/\Omega_{\Sc'}}\delta_{\Sc_i',1}\idop_{4}\prod_{j\in \Omega_{\Sc'}}\frac{1-\delta_{\Sc_j',1}}{3}\bm{\sigma}_j\cdot\bar{\bsigma}_j,
\end{dmath}
where in the second equality the sum is over the powerset of $\{1,2,\cdots,N\}$, i.e, all the ($2^N$) subsets of $\{1,2,\cdots,N\}$.
The sum above essentially contains every possible string of the form $\Sc\otimes \Sc$, i.e the same operator appears on both copies of the system.
Note that for a given string $\Sc'$, there is only one nonzero term in the sum. For each site $i$, we either put an $\idop_4$ if $\Sc_i'=\idop$ or we place $\frac{1}{3}\bsigma_i\cdot\bar{\bsigma}_i$, if $\Sc_i'$ is any other generator.
The set $\Omega_{\Sc'}$ therefore represents the support of the string $\Sc'$.

Before proceeding, let us summarize the high-level idea behind the derivation that follows.
Our tasks consist of the following: 
\begin{enumerate}
	\item First, we need to apply the global interaction $\Uzz\otimes \Uzz$ on \cref{eq:Haar-qubits-general}, giving us $\Ic_2$. 
	\item Then, we need to apply the layer of single-site Haar unitaries $\Uhar\otimes \Uhar$, and average over the Haar distribution on each site, giving us $\Ic_3$. 
	\item Finally, we need to multiply the result by $\Sc\otimes \Sc$ and take the trace, giving us $W_{\Sc,\Sc'}$.
\end{enumerate}
Recall that 
\begin{equation}
\label{eq:Uzz-def}
\Uzz = e^{-i\frac{g'}{2}\sum_{i<j}Z_iZ_j},
\end{equation}
where in the main text we have assumed $g'=\frac{g}{\sqrt{N}}$.
To perform the first step, we will make use of the formulas
\begin{align}
\label{eq:rotation-X-formula}
\Uzz X_r \Uzz^\dagger &= X_r\cos(g'\sum_{i\neq r}Z_i)+Y_r\sin(g'\sum_{i\neq r}Z_i),\\
\label{eq:rotation-Y-formula}
\Uzz Y_r \Uzz^\dagger &= Y_r\cos(g'\sum_{i\neq r}Z_i)-X_r\sin(g'\sum_{i\neq r}Z_i).
\end{align}
Now, note that each term in the sum in \cref{eq:Haar-qubits-general} is a product of single-site operators. By performing our first task, using \cref{eq:rotation-X-formula,eq:rotation-Y-formula}, we will obtain complicated operators, like those appearing on the right-hand-side of \cref{eq:rotation-X-formula,eq:rotation-Y-formula}, that are supported on a large number of sites.
In order to perform the second step, we can make use of \cref{eq:Haar-average-formula}.
However, to use \cref{eq:Haar-average-formula}, we need $A$ to be a single-site operator. Thus, we will have to break down complicated operators, like those appearing on the right-hand-side of \cref{eq:rotation-X-formula,eq:rotation-Y-formula}, into sums of simple terms consisting of products of single-site operators.
This will allow us to use \cref{eq:Haar-average-formula}, after which we can easily perform the last step, 3, since this will only require taking traces of single-site operators.

The result of step 1 and 2 can be simplified by noting that \cref{eq:Haar-qubits-general} contains all possible strings of the form $\Sc\otimes \Sc$. Hence, it is instructive to first consider the result of applying $\Uzz\otimes \Uzz$ and $\Uhar\otimes\Uhar$ to a single string of this form. 
Note that the result of applying $\Uzz\otimes \Uzz$, $\Uhar\otimes\Uhar$, and averaging over the Haar unitaries is invariant if we replace any number of $X$s in the string by $Y$s or vice-versa. To see this, we use the fact that we can change a $X$ into a $Y$ (or vice-versa) by applying a rotation about the $Z$ axis, i.e $e^{-i\frac{\pi}{4}Z}Xe^{i\frac{\pi}{4}Z}=Y$. This rotation clearly commutes with $\Uzz$ and can be absorbed into $\Uhar$, since by definition, the Haar measure is invariant under multiplication by any unitary.

This means that we may calculate the result for a single representative string from each group and multiply by the degeneracy.
Let us denote $\Omega_{\Sc}$ the support of some string $\Sc$.
We can further divide $\Omega_{\Sc}$ based on the number and location of $Z$s in the string. 
Define the subset $\Sigma \subseteq \Omega$ as the set of all sites with $Z$ in them, and the remaining sites (with either $X$s or $Y$s) by $\Lambda =\Omega \setminus \Sigma$. 
For strings that are supported on $k$ sites (i.e $\abs{\Omega_{\Sc}}=k$), with fixed number and position of $Z$s, the degeneracy is $2^{\abs{\Lambda}}$.

Without loss of generality, we can therefore consider strings composed of either $X$s or $Z$s.
Consider the string $\prod_{i \in \Lambda}X_i\bar{X}_i\prod_{j\in \Sigma}Z_j\bar{Z}_j$.
To apply $\Uzz$, we can use the fact that $\comm{X_iX_j}{Z_iZ_j}=0$.
We get
\begin{dmath}
	(\Uzz\otimes \Uzz) (\prod_{i \in \Lambda}X_i\bar{X}_i\prod_{j\in \Sigma}Z_j\bar{Z}_j)(\Uzz\otimes \Uzz)^\dagger
	=
	\prod_{i\in \Lambda}\qty[\qty(X_i\cos(Q_\Lambda)+Y_i\sin(Q_\Lambda))\qty(\bar{X}_i\cos(\bar{Q}_\Lambda)+\bar{Y}_i\sin(\bar{Q}_\Lambda))]\prod_{j\in \Sigma}Z_j\bar{Z}_j
\end{dmath}
where we used \cref{eq:rotation-X-formula}. Here, $Q_\Lambda$ acts on all sites except those in $\Lambda$, i.e $Q_\Lambda\equiv g'\sum_{l\notin \Lambda}Z_l$. 

We see that we can safely apply the Haar unitaries and perform the Haar average on sites inside of $\Lambda$, since all the cosines and sines and the $Z\bar{Z}$ act on sites outside of $\Lambda$. 
With slight abuse of notation, let us denote $\expval{(\Uhar\otimes \Uhar)A(\Uhar\otimes \Uhar)^\dagger}$ by simply $\expval{A}$ where it is understood that the Haar unitaries act only on the support of $A$.

From \cref{eq:Haar-average-formula}, one can easily check that $\expval{X_i\bar{Y}_i}=0$, so the cross terms in the above expression will vanish. Only $\expval{X_i\bar{X}_i}=\expval{Y_i\bar{Y}_i}\equiv V_i$ will remain. Here the single site operator $V_i$ is defined as $V_j = -\frac{1}{3}\idop_4+\frac{2}{3}F$.
Explicitly, we find
\begin{dmath}
	\expval{(\Uzz\otimes \Uzz) (\prod_{i \in \Lambda}X_i\bar{X}_i\prod_{j\in \Sigma}Z_j\bar{Z}_j)(\Uzz\otimes \Uzz)^\dagger}
	=
	\prod_{i\in \Lambda}V_i\expval{\cos[\abs{\Lambda}](R_\Lambda)\prod_{j\in \Sigma}Z_j\bar{Z}_j}
\end{dmath}
where $R_\Lambda = \bar{Q}_\Lambda-Q_\Lambda$.

Combining this with the discussion above, we find that $\Ic_3$ may be written as 
\begin{dmath}
	\label{eq:Haar-Twist-Haar-complete}
	\Ic_3=\delta_{\Sc',\Sc''}\sum_{\Omega_{\Sc'} \subset \{1,2,\cdots,N\}}\qty(\prod_{j\notin\Omega_{\Sc'}}\delta_{\Sc_j',1})\qty(\prod_{i\in\Omega_{\Sc'}}\frac{1-\delta_{\Sc_i',1}}{3})\sum_{\Lambda \subset \Omega_{\Sc'}}2^{\abs{\Lambda}}\qty(\prod_{m\in \Lambda}V_m)\expval{\cos[\abs{\Lambda}](R_\Lambda)\prod_{n\in \Omega_{\Sc'}\setminus \Lambda}Z_n\bar{Z}_n}.
\end{dmath}

It remains to compute $\expval{\cos[\abs{\Lambda}](R_\Lambda)\prod_{n\in \Omega_{\Sc'}\setminus \Lambda}Z_n\bar{Z}_n}$.
To do so we expand the cosine as follows $\cos[k](x) =
\frac{1}{2^k}\sum_{n=0}^k\mqty(k\\n)\cos[(2n-k)x]$,
\begin{dmath}
	\label{eq:cos[Lambda]-prod-ZZ}
	\expval{\cos[\abs{\Lambda}](R_\Lambda)\prod_{n\in \Omega_{\Sc'}\setminus \Lambda}Z_n\bar{Z}_n}=
	\frac{1}{2^{\abs{\Lambda}}}\sum_{l=0}^{\abs{\Lambda}}\mqty(\abs{\Lambda}\\l)\expval{\cos((2l-\abs{\Lambda})R_\Lambda)\prod_{n\in \Omega_{\Sc'}\setminus \Lambda}Z_n\bar{Z}_n}.
\end{dmath}
To proceed we can pull a single-site operator out of $R_\Lambda$. Since $R_\Lambda = \sum_{k\notin  \Lambda} D_k$ where $D_k=g'(\bar{Z}_k-Z_k)$, we can pull out a $D_j\qc j\in \Omega_{\Sc'}\setminus\Lambda$ so that $R_\Lambda = R_{\Lambda\cup\{j\}}+D_j$. We then use the trig identity
\begin{equation}
\label{eq:trig-plus-identity}
\cos((2l-\abs{\Lambda})R_\Lambda)= \cos((2l-\abs{\Lambda})R_{\Lambda\cup\{j\}})\cos((2l-\abs{\Lambda})D_j)-\sin((2l-\abs{\Lambda})R_{\Lambda\cup\{j\}})\sin((2l-\abs{\Lambda})D_j).
\end{equation}
This allows us to perform the Haar average over site $j$. The sine term will not contribute, since $\expval{\sin((2l-\abs{\Lambda})D_j)Z_j\bar{Z}_j}=0$.
Repeating this procedure recursively for all sites in $\Omega_{\Sc'}\setminus\Lambda$, we get
\begin{dmath}
	\expval{\cos((2l-\abs{\Lambda})R_\Lambda)\prod_{n\in \Omega_{\Sc'}\setminus \Lambda}Z_n\bar{Z}_n}=
	\expval{\cos((2l-\abs{\Lambda})R_{\Omega_{\Sc'}})}\prod_{n\in \Omega_{\Sc'}\setminus \Lambda}\expval{\cos((2l-\abs{\Lambda})D_n)Z_n\bar{Z}_n}.
\end{dmath}
Continuing the procedure for the $\expval{\cos((2l-\abs{\Lambda})R_{\Omega_{\Sc'}})}$ term, we have
\begin{dmath}
	\expval{\cos((2n-\abs{\Lambda})R_\Lambda)\prod_{n\in \Omega_{\Sc'}\setminus \Lambda}Z_n\bar{Z}_n}=\prod_{t\notin \Omega_{\Sc'}}\expval{\cos((2l-\abs{\Lambda})D_t)}
	\prod_{n\in \Omega_{\Sc'} \setminus \Lambda}\expval{\cos((2l-\abs{\Lambda})D_n)Z_n\bar{Z}_n}.
\end{dmath}
Using $\cos((2l-\abs{\Lambda})D)=\cos[2]((2l-\abs{\Lambda})g')+Z\bar{Z}\sin[2]((2l-\abs{\Lambda})g')$ gives
\begin{dmath}
	\expval{\cos((2n-\abs{\Lambda})R_\Lambda)\prod_{n\in \Omega_{\Sc'}\setminus \Lambda}Z_n\bar{Z}_n}=\prod_{t\notin \Omega_{\Sc'}}\qty(\cos[2]((2l-\abs{\Lambda})g')+V_t\sin[2]((2l-\abs{\Lambda})g'))
	\prod_{n\in \Omega_{\Sc'} \setminus \Lambda}\qty(\cos[2]((2l-\abs{\Lambda})g')V_n+\sin[2]((2l-\abs{\Lambda})g')).
\end{dmath}
Putting things together, we find that \cref{eq:cos[Lambda]-prod-ZZ} is
\begin{dmath}
	\expval{\cos[\abs{\Lambda}](R_\Lambda)\prod_{n\in \Omega_{\Sc'}\setminus \Lambda}Z_n\bar{Z}_n}=\frac{1}{2^{\abs{\Lambda}}}\sum_{l=0}^{\abs{\Lambda}}\mqty(\abs{\Lambda}\\l)\prod_{t\notin \Omega_{\Sc'}}\qty(\cos[2]((2l-\abs{\Lambda})g')+V_t\sin[2]((2l-\abs{\Lambda})g'))
	\prod_{n\in \Omega_{\Sc'} \setminus \Lambda}\qty(\cos[2]((2l-\abs{\Lambda})g')V_n+\sin[2]((2l-\abs{\Lambda})g')),
\end{dmath}
and finally, $\Ic_3$ is given by
\begin{multline}	
\label{eq:Haar-Twist-Haar-final}
\Ic_3=\delta_{\Sc',\Sc''}
\sum_{\Omega_{\Sc'} \subset \{1,2,\cdots,N\}}\qty(\prod_{j\notin\Omega_{\Sc'}}\delta_{\Sc_j',1})\qty(\prod_{i\in\Omega_{\Sc'}}\frac{1-\delta_{\Sc_i',1}}{3})\sum_{\Lambda \subset \Omega_{\Sc'}}\qty(\prod_{m\in \Lambda}V_m)\sum_{l=0}^{\abs{\Lambda}}\mqty(\abs{\Lambda}\\l)\\
\times\prod_{t\notin \Omega_{\Sc'}}\qty(\cos[2]((2l-\abs{\Lambda})g')+V_t\sin[2]((2l-\abs{\Lambda})g'))
\prod_{n\in \Omega_{\Sc'} \setminus \Lambda}\qty(\cos[2]((2l-\abs{\Lambda})g')V_n+\sin[2]((2l-\abs{\Lambda})g')).
\end{multline}

To compute the $W_{\Sc,\Sc'}$-matrix from \cref{eq:a_S_2-haar-general}, it remains to take the trace of \cref{eq:Haar-Twist-Haar-final} with $\Sc\otimes \Sc$ and divide by $2^{2N}$, i.e

\begin{equation}
W_{\Sc,\Sc'} = \frac{1}{2^{2N}}\tr(\Ic_3(\Sc\otimes\Sc))
\end{equation}
Using \cref{eq:Haar-Twist-Haar-final} together with $\tr(V_i (\Sc_i\otimes\Sc_i))=\frac{4}{3}(1-\delta_{\Sc_i,1})$, gives

\begin{multline}	
\label{eq:W-Haar-Twist-Haar-initial}
W_{\Sc,\Sc'} =
\frac{1}{2^{2N}}\sum_{\Omega_{\Sc'} \subset \{1,2,\cdots,N\}}\qty(\prod_{j\notin\Omega_{\Sc'}}\delta_{\Sc_j',1})\qty(\prod_{i\in\Omega_{\Sc'}}\frac{1-\delta_{\Sc_i',1}}{3})\sum_{\Lambda \subset \Omega_{\Sc'}}\qty(\prod_{m\in \Lambda}\frac{4}{3}(1-\delta_{\Sc_m,1}))\sum_{l=0}^{\abs{\Lambda}}\mqty(\abs{\Lambda}\\l)\\
\times\prod_{t\notin \Omega_{\Sc'}}\qty(\cos[2]((2l-\abs{\Lambda})g')4\delta_{\Sc_t,1}+\frac{4}{3}(1-\delta_{\Sc_t,1})\sin[2]((2l-\abs{\Lambda})g'))\\
\times
\prod_{n\in \Omega_{\Sc'} \setminus \Lambda}\qty(\cos[2]((2l-\abs{\Lambda})g')\frac{4}{3}(1-\delta_{\Sc_n,1})+4\delta_{\Sc_n,1}\sin[2]((2l-\abs{\Lambda})g')).
\end{multline}
Note that because of $\prod_{m\in \Lambda}\frac{4}{3}(1-\delta_{\Sc_m,1})$ in \cref{eq:W-Haar-Twist-Haar-initial}, $\Lambda$ is constrained to be in $\Omega_\Sc\cap\Omega_{\Sc'}$. 
The matrix elements of $W$ are
\begin{multline}	
\label{eq:W-Haar-Twist-Haar-elem}
W = \frac{1}{2^{2N}}\qty(\frac{1}{3})^{\abs{\Omega_{\Sc'}}}\sum_{\Lambda \subset \Omega_\Sc\cap\Omega_{\Sc'}}\qty(\frac{4}{3})^{\abs{\Lambda}}\biggl[\sum_{l=0,2l\neq \abs{\Lambda}}^{\abs{\Lambda}}\mqty(\abs{\Lambda}\\l)
\qty(4\cos[2]((2l-\abs{\Lambda})g'))^{N-\abs{\Omega_{\Sc}\cup\Omega_{\Sc'}}}\\
\times
\qty(\frac{4}{3}\sin[2]((2l-\abs{\Lambda})g'))^{\abs{\Omega_{\Sc}\setminus\Omega_{\Sc'}}}
\times
\qty(\frac{4}{3}\cos[2]((2l-\abs{\Lambda})g'))^{\abs{\Omega_{\Sc}\cap\Omega_{\Sc'}}-\abs{\Lambda}}
\times
\qty(4\sin[2]((2l-\abs{\Lambda})g'))^{\abs{\Omega_{\Sc'}\setminus\Omega_{\Sc}}}\\
+\delta_{2l,\abs{\Lambda}}\mqty(\abs{\Lambda}\\\abs{\Lambda}/2)\prod_{t\notin \Omega_{\Sc'}}\qty(4\delta_{\Sc_t,1})\prod_{n\in \Omega_{\Sc'} \setminus \Lambda}\qty(\frac{4}{3}(1-\delta_{\Sc_n,1}))
\biggr].
\end{multline}
Note that the last term is only nonzero when both $2l=\abs{\Lambda}$ and $\Omega_\Sc= \Omega_{\Sc'}$. The last condition is equivalent to $\abs{\Omega_{\Sc}}+\abs{\Omega_{\Sc'}}-2\abs{\Omega_{\Sc}\cap\Omega_{\Sc'}}=0$.

We can combine all constant factors (with the same result holding for the $2l=\abs{\Lambda}$ term)
\begin{dmath}
	\frac{1}{2^{2N}}\qty(\frac{1}{3})^{\abs{\Omega_{\Sc'}}}\qty(\frac{4}{3})^{\abs{\Lambda}}4^{N-\abs{\Omega_{\Sc}\cup\Omega_{\Sc'}}}\qty(\frac{4}{3})^{\abs{\Omega_{\Sc}\setminus\Omega_{\Sc'}}}\qty(\frac{4}{3})^{\abs{\Omega_{\Sc}\cap\Omega_{\Sc'}}-\abs{\Lambda}}4^{\abs{\Omega_{\Sc'}\setminus\Omega_{\Sc}}}
	=
	\qty(\frac{1}{3})^{\abs{\Omega_{\Sc'}}+\abs{\Omega_{\Sc}}}.
\end{dmath}

Now, note that $\Lambda$ only appears in \cref{eq:W-Haar-Twist-Haar-elem} as $\abs{\Lambda}$. Thus, we can replace the sum over subsets of $\Omega_{\Sc}\cap\Omega_{\Sc'}$ as $\sum_{\Lambda \subset \Omega_\Sc\cap\Omega_{\Sc'}} = \sum_{k=0}^{\abs{\Omega_{\Sc}\cap\Omega_{\Sc'}}}\mqty(\abs{\Omega_{\Sc}\cap\Omega_{\Sc'}}\\k)$.
Thus, the $W$ matrix can be written as 
\begin{align}
\label{eq:W-Haar-Twist-Haar-final}
W_{\Sc,\Sc'} &= W(\abs{\Omega_{\Sc}},\abs{\Omega_{\Sc'}},\abs{\Omega_{\Sc}\cap\Omega_{\Sc'}})\\\nonumber &= 
\qty(\frac{1}{3})^{\abs{\Omega_{\Sc'}}+\abs{\Omega_{\Sc}}}\sum_{k=0}^{\abs{\Omega_{\Sc}\cap\Omega_{\Sc'}}}\mqty(\abs{\Omega_{\Sc}\cap\Omega_{\Sc'}}\\k)\biggl[\sum_{l=0,2l\neq k}^{k}\mqty(k\\l)\qty[\cos[2]((2l-k)g')]^{N-k-(\abs{\Omega_{\Sc}}+\abs{\Omega_{\Sc'}}-2\abs{\Omega_{\Sc}\cap\Omega_{\Sc'}})}\\ \nonumber
&\times
\qty[\sin[2]((2l-k)g')]^{\abs{\Omega_{\Sc}}+\abs{\Omega_{\Sc'}}-2\abs{\Omega_{\Sc}\cap\Omega_{\Sc'}}}+\delta_{2l,k}\delta_{\abs{\Omega_{\Sc}}+\abs{\Omega_{\Sc'}}-2\abs{\Omega_{\Sc}\cap\Omega_{\Sc'}},0}\mqty(k\\k/2)\biggr],
\end{align}
which is what appears in Eq.~(5) of the main text, with the identification $w=\abs{\Omega_{\Sc}},w'=\abs{\Omega_{\Sc'}},v=\abs{\Omega_{\Sc}\cap\Omega_{\Sc'}}$. In the main text, we also dropped the $\delta_{2l,k}\delta_{\abs{\Omega_{\Sc}}+\abs{\Omega_{\Sc'}}-2\abs{\Omega_{\Sc}\cap\Omega_{\Sc'}},0}$ term and the $2l\neq k$ restriction in the sum which requires one to be careful to identify $0^0$ as 1.
From this expression it is clear that $W$ is a real symmetric ($W_{\Sc,\Sc'}=W_{\Sc',\Sc}$) matrix with all positive matrix elements.

\section{Master equation for simple initial operator}
\label{sec:master-eq}

Let us now assume that the initial operator $\Oc$ starts as a single-site operator on site $1$ without loss of generality.
We may further assume that we start with $X_1$, i.e $a_{\Sc} = \delta_{\Sc,X_1}$. 
Since the circuit will involve random Haar unitaries, let us consider the result of applying a Haar random unitary on $X_1$, which, after averaging over the Haar unitary, will be $\frac{1}{3}(X_1+Y_1+Z_1)$, which already does not contain any information about the specific generator we picked. 
Let us therefore pick this as the initial conditions at $t=0$ for the master equation, \cref{eq:a_S_2-haar-general}, 
\begin{equation}
\label{eq:master-eq-initial-cond}
\expval{a_\Sc^2(t=0)} = \begin{cases}
\frac{1}{3}\qif \Sc = X_1,Y_1,Z_1,\\
0 \qotherwise.
\end{cases}
\end{equation}
We now claim that for these initial conditions, the probabilities $\expval{a_\Sc^2(t)}$ only depend on the string weight $w\equiv \abs{\Omega_{\Sc}}$ and the weight on site $1$, $w_1\equiv \abs{\Omega_{\Sc}\cap\{1\}}$. Note that $w_1$ takes values either $0$ or $1$.
In light of this, it is convenient to account for the number of string configurations with constant $w$ and $w_1$ by defining the operator weight probability $h_t$,
\begin{dmath}
	\label{eq:h-def-norm}
	h_t(w,w_1) = \expval{a_\Sc^2(t)} D(w,w_1),
\end{dmath} 
where $D(w,w_1)$ is the number of string configurations for a given $w$ and $w_1$.  Since $\sum_{\Sc'} = \sum_{w_1=0,1}\sum_{w=w_1}^{N-1+w_1}3^k\mqty(N-1\\w-w_1)$, we have 
\begin{dmath}
	D(w,w_1) = 3^w\mqty(N-1\\w-w_1).
\end{dmath}
Note that $h_t(w,w_1)$ is a valid (normalized) probability distribution since $\sum_{w_1=0,1}\sum_{w=w_1}^{N-1+w_1}h_t(w,w_1)= \sum_{w_1=0,1}\sum_{w=w_1}^{N-1+w_1} \expval{a_\Sc^2(t)} D(w,w_1) = \sum_{\Sc}  \expval{a_\Sc^2(t)}=1$, using the fact that $a_{\Sc}^2$ are probabilities that sum to 1. Thus, $h_t(w,w_1)$ gives the probability of $\Oc(t)$ being a string of total weight $w$ with a weight of $w_1$ on the initial site 1.

The claim above can be proved by induction. The base case is trivial to see, by multiplying the initial conditions \cref{eq:master-eq-initial-cond} by the transition matrix $W$ from \cref{eq:W-Haar-Twist-Haar-final} (see also \cref{sec:onestep}).
The inductive step proceeds as follows. 
First, we decompose the sum over strings $\Sc'$ as $\sum_{\Sc'} = \sum_{\Omega_{\Sc'}\subset \{1,\cdots,N\}}3^{\abs{\Omega_{\Sc'}}}$, which yields
\begin{dmath}
	\label{eq:master-eq-h-superset-sum-initial}
	\expval{a_\Sc^2(t+1)} = \sum_{\Omega_{\Sc'}\subset \{1,\cdots,N\}}\frac{1}{D(\abs{\Omega_{\Sc'}},\abs{\Omega_{\Sc'}\cap{1}})}3^{\abs{\Omega_{\Sc'}}}W(\abs{\Omega_{\Sc}},\abs{\Omega_{\Sc'}},\abs{\Omega_{\Sc}\cap\Omega_{\Sc'}})h_{t}(\abs{\Omega_{\Sc'}},\abs{\Omega_{\Sc'}\cap{1}}).
\end{dmath}
We then split the sum over terms where $\abs{\Omega_{\Sc'}\cap\{1\}}=0$ or  $\abs{\Omega_{\Sc'}\cap\{1\}}=1$. 
For each of these terms, we further decompose the sum over terms with equal $\abs{\Omega_{\Sc'}}$. The remaining sum can be written as a sum over different values of the overlap $\abs{\Omega_{\Sc}\cap\Omega_{\Sc'}}$.
The final result is 

\begin{dmath}
	\expval{a_\Sc^2(t+1)} = \sum_{k=0}^{N-1}3^k\qty[		\sum_{m=0}^{\min\{\abs{\Omega_{\Sc}}-\abs{\Omega_{\Sc}\cap\{1\}},k\}}\mqty(\abs{\Omega_{\Sc}}-\abs{\Omega_{\Sc}\cap\{1\}}\\m)\mqty(N-1-\abs{\Omega_{\Sc}}+\abs{\Omega_{\Sc}\cap\{1\}}\\k-m)W(\abs{\Omega_{\Sc}},k,m)]\frac{h_t(k,0)}{D(k,0)}
	+
	\sum_{k=1}^{N}3^k\qty[\sum_{m=\abs{\Omega_{\Sc}\cap\{1\}}}^{\min\{\abs{\Omega_{\Sc}},k-1+\abs{\Omega_{\Sc}\cap\{1\}}\}}\mqty(\abs{\Omega_{\Sc}}-\abs{\Omega_{\Sc}\cap\{1\}}\\ m-\abs{\Omega_{\Sc}\cap\{1\}})\mqty(N-1+\abs{\Omega_{\Sc}\cap\{1\}}-\abs{\Omega_{\Sc}}\\k-m-1+\abs{\Omega_{\Sc}\cap\{1\}})W(\abs{\Omega_{\Sc}},k,m)]\frac{h_t(k,1)}{D(k,1)}.
\end{dmath}
Here, the first binomial in each bracket counts the number of ways one can choose the part of 
$\Omega_{\Sc'}$ that is overlapping with $\Omega_{\Sc}$ and the second binomial counts the number of ways to choose the non-overlapping part of $\Omega_{\Sc'}$.
It is clear at this point that the right-hand-side is a function of $w=\abs{\Omega_{\Sc}}$ and
$w_1=\abs{\Omega_{\Sc}\cap{1}}$.
Thus, replacing $\expval{a_\Sc^2(t+1)}$ by \cref{eq:h-def-norm} and simplifying gives
\begin{dmath}
	\label{eq:htilde-master-eq}
	h_{t+1}(w,w_1) = \sum_{w_1'=0,1}\sum_{w'=w_1'}^{N-1+w_1'}\mathcal{R}(w,w_1,w',w_1')h_{t}(w',w_1')\end{dmath}
where the $2N\times 2N$ matrix $\mathcal{R}$ is 
\begin{align}
\label{eq:tildeW-def}
\mathcal{R}(w,w_1,w',w_1') = 
3^{w}\sum_{m=\max\{0,w+w'-N+1-w_1-w_1'\}}^{\min\{w-w_1,w'-w_1'\}}\mqty(w'-w_1'\\m)\mqty(N-1-w'+w_1'\\w-w_1-m) W(w,w',m+w_1w_1'),
\end{align}
where $w_1,w_1' \in \{0,1\},w\in [w_1,N-1+w_1],w'\in [w_1',N-1+w_1']$, and for completeness 
\begin{align}
\label{eq:W-def}
W(w,w',v) &=  
\qty(\frac{1}{3})^{w+w'}\sum_{k=0}^{v}\mqty(v\\k)\biggl[\sum_{l=0,2l\neq k}^{k}\mqty(k\\l)\qty[\cos[2]((2l-k)g')]^{N-k-(w+w'-2v)}\\ \nonumber
&\times
\qty[\sin[2]((2l-k)g')]^{w+w'-2v}+\delta_{2l,k}\delta_{w+w'-2v,0}\mqty(k\\k/2)\biggr].
\end{align}
One may verify that $\sum_{i}\mathcal{R}_{i,j}=1$ where $i=(w,w_1)$ and $j=(w',w_1')$. This means that if we start with normalized $h_0$, we will have a valid (normalized) probability distribution at later times.

The initial conditions become
\begin{dmath}
	\label{eq:init-cond-full}
	h_0(w,w_1)=\begin{cases}
		1\qif w =w_1=1,\\
		0 \qotherwise.
	\end{cases}
\end{dmath}
To get the probability of having a specific weight, we can sum over $w_1$,
\begin{dmath}
	\label{eq:h-total-def}
	h(w)=\begin{cases}
		h(0,0)\qif w =0,\\
		h(N,1)\qif w =N,\\
		h(w,0)+h(w,1) \qotherwise.
	\end{cases}
\end{dmath}
Note that $h(0,0)$ does not actually participate in the dynamics since $\mathcal{R}(0,0,w',w_1')=W(0,w',0)=\delta_{w',0}$.

\section{Continuum approximation}
\label{sec:cont-approx}
We assume here the normalization $g'=\frac{g}{\sqrt{N}}$. 
The first step is to approximate $W(w,w',v)$ for small $g$.
We consider the two cases $w+w'-2v = 0,1$ which amount to a change of the string weight by $0,\pm1$ and give rise to terms up to $g^2$.

Taylor expanding the factors of cosine and sine appearing in \cref{eq:W-def}, up to $g^2$, gives
\begin{dmath}
	\qty[\cos[2]((2l-k)\frac{g}{\sqrt{N}})]^{N-k-(w+w'-2v)}\qty[\sin[2]((2l-k)\frac{g}{\sqrt{N}})]^{w+w'-2v} \approx \begin{cases}
		\frac{g^2 (k-2 l)^2 (k-N)}{N}+1\qif w+w'-2v=0,\\
		\frac{g^2 (k-2 l)^2}{N}\qif w+w'-2v=1.
	\end{cases}
\end{dmath}
In general, the $w+w'-2v = n,\,n\in \mathbb{N}_{>0}$ case will scale as  $O(g^{2n})$.
We can now perform the sums over $l$ and $k$ appearing in \cref{eq:W-def}. We find
\begin{dmath}
	\label{eq:W-approx}
	W(w,w',v) \approx \qty(\frac{1}{3})^{w+w'-v}\begin{cases}
		1+g^2\frac{2v}{3^2N}(1-3N+2v)\qif w+w'-2v=0,\\
		g^2\frac{2v}{3N}\qif w+w'-2v=1.
	\end{cases}
\end{dmath}
The higher order terms will scale at most like $O(g^4 N^2/N^2)=O(g^4)$ and so for small $g$, the above expression for $W(w,w',v)$ is an excellent approximation. 
In the general case of $g'= \frac{g}{N^a},\,a\ge0$, the above Taylor expansion yields a series expression for $W(w,w',v)$ where the $n$th term scales at most as $O(g^{2n}N^n/N^{2na})$. Thus, for $a<\frac{1}{2}$, the series is not convergent, and \cref{eq:W-approx} does not constitute a good approximation. 
Below, we assume $a=\frac{1}{2}$, but all results and expressions in this section are applicable for $a\ge \frac{1}{2}$ as well, with the appropriate replacement of $g$. For some discussion of the $a=0$ case, see \cref{sec:onestep}.

Let us now consider the $\mathcal{R}$ matrix. The $w+w'-2v = 0,1$ cases contribute to the diagonal as well as super- and sub-diagonals of each block of $\mathcal{R}$.
These matrix elements are 
\begin{align}
\mathcal{R}(w,0,w',0) &= \delta_{w,w'}3^wW(w,w',w')
+\delta_{w,w'+1}3^w(N-w'-1)W(w,w',w')\\ \nonumber
&+\delta_{w,w'-1}3^ww'W(w,w',w'-1)
+O(g^4),\\
\mathcal{R}(w,1,w',0) &= 
\delta_{w,w'+1}3^wW(w,w',w')
+O(g^4),\\
\mathcal{R}(w,0,w',1) &= 
\delta_{w,w'-1}3^wW(w,w',w'-1)
+O(g^4),\\
\mathcal{R}(w,1,w',1) &= 
\delta_{w,w'}3^wW(w,w',w')
+\delta_{w,w'+1}3^w(N-w')W(w,w',w')\\ \nonumber
&+\delta_{w,w'-1}3^w(w'-1)W(w,w',w'-1)
+O(g^4).
\end{align}
Writing out the master equation, \cref{eq:htilde-master-eq}, within the $g^2$ approximation, we have two coupled equations for the two ($w_1=0,1$) blocks:
\begin{align}
\label{eq:h-discrete-approx-g2-1}
\frac{h_{t+1}(w,0)-h_t(w,0)}{g^2} =& \frac{2w}{9N}h_t(w+1,1) +\frac{2w}{9N}(1-3N+2w)h_t(w,0)\\ \nonumber
+&\frac{2(N-w)}{3N}(w-1)h_t(w-1,0)
+\frac{2w(w+1)}{9N}h_t(w+1,0),\\
\label{eq:h-discrete-approx-g2-2}
\frac{h_{t+1}(w,1)-h_t(w,1)}{g^2} =& \frac{2(w-1)}{3N}h_t(w-1,0) +\frac{2w}{9N}(1-3N+2w)h_t(w,1)\\ \nonumber
+&\frac{2(w-1)}{3N}(N-w+1)h_t(w-1,1)
+\frac{2w^2}{9N}h_t(w+1,1).
\end{align}
Note that the coupling between the two $w_1$ sectors scales as $w/N$. Since the initial conditions are constrained to the $w_1=1$ sector [see \cref{eq:init-cond-full}], the early time dynamics will remain approximately in $h_t(w,1)$ (i.e $h_t(w,0)\approx 0$ at early times) until $w$ reaches $O(N)$.

By adding \cref{eq:h-discrete-approx-g2-1,eq:h-discrete-approx-g2-2}, we get a closed equation for the total operator weight probability $h_t(w) \equiv h_t(w,0)+h_t(w,1)$ 
\begin{dmath}
	\frac{h_{t+1}(w)-h_t(w)}{g^2} = \frac{2w(w+1)}{9N}h_t(w+1) +\frac{2w}{9N}(1-3N+2w)h_t(w)
	+\frac{N-w+1}{3N}2(w-1)h_t(w-1).
\end{dmath}
Up to now, the only approximation we made was the expansion up to $g^2$.
We now assume that $h(w,t)$ varies slowly with respect to $g^2t$ and $w$, and replace finite differences by derivatives which yields a Fokker-Planck equation
\begin{equation}
\label{eq:Fokker-Planck}
\partial_\tau h(w,\tau) = -\partial_w\qty(D_1(w)h(w,\tau))
+\partial_w^2\qty(D_2(w)h(w,\tau)),
\end{equation}
where we introduced a rescaled time $\tau=g^2t$. Note that \cref{eq:h-discrete-approx-g2-1,eq:h-discrete-approx-g2-2} individually are not in the form of a Fokker-Planck equation, but their sum is.
The drift and diffusion coefficients are
\begin{align}
D_1(w) &= \frac{2(4+w+3Nw-4w^2)}{9N},\\
D_2(w) &= \frac{-3+3N(w-1)+7w-2w^2}{9N}.
\end{align}
In terms of the scaled weight $\phi \equiv w/N$, the Fokker-Planck equation takes the form
\begin{equation}
\label{eq:Fokker-Planck-phi}
\partial_\tau h(\phi,\tau) = -\partial_\phi\qty(\frac{2}{3}\qty(\phi-\frac{4}{3}\phi^2)h(\phi,\tau))
+\partial_\phi^2\qty(\qty(\frac{\phi}{3N}-\frac{2}{9}\frac{\phi^2}{N})h(\phi,\tau)),
\end{equation}
where we dropped all the $O(1/N)$ terms from the drift coefficient and all the $O(1/N^2)$ terms from the diffusion.

\section{Relation between the average of the squared commutator and the mean operator weight}
\label{sec:squared-comm-w}
In this section, we derive the relation between the average of the squared commutator, defined in Eq.~(1) of the main text, and the operator weight probability $h_t(w,w_1)$.

Let us start with Eq.~(1) of the main text, and, without loss of generality, pick the two operators to be $X_1$ at position 1 and $Y_r$ at position $r>1$
\begin{equation}
\label{eq:sqa-com-def}
\Cc(r,t)= -\frac{1}{2}\tr(\rho_\infty\comm{X_1(t)}{Y_r}^2),
\end{equation} 
where $\rho_{\infty}$ is the infinite-temperature Gibbs state, and $X_1(t)$ is the Heisenberg evolved operator.
Using \cref{eq:O-heisenberg-expansion-strings}, the commutator in \cref{eq:sqa-com-def} can be written as
\begin{equation}
\comm{X_1(t)}{Y_r}^2=\qty(\sum_\Sc a_\Sc(t)\comm{\Sc}{Y_r})^2=
\qty(2\sum_{\Sc:\Sc_r=X,Z}a_\Sc(t)\Sc Y_r)^2,
\end{equation}
which gives 
\begin{align}
\label{eq:Cc-simplified}
\Cc(r,t) =& -2\sum_{\Sc:\Sc_r=X,Z}\sum_{\Sc':\Sc'_r=X,Z}a_\Sc(t)a_{\Sc'}(t)\tr(\rho_\infty\Sc Y_r\Sc'Y_r)\\
=&
2\sum_{\Sc:\Sc_r=X,Z}a_\Sc(t)^2,
\end{align}
where we used $\tr(\rho_\infty\Sc\Sc')=\delta_{\Sc\Sc'}$ and the fact that different Pauli matrices anti-commute.
Here the sum is constrained to be over all strings that have an $X$ or a $Z$ on site $r$.

The average of \cref{eq:Cc-simplified} over many realizations of the random circuit is therefore given by
\begin{equation}
\label{eq:average-c}
\expval{\Cc(r,t)} = 2\sum_{\Sc:\Sc_r=X,Z}\expval{a_\Sc(t)^2},
\end{equation}
where the evolution of $\expval{a_{\Sc}^2(t)}$ is what we calculated in the previous sections.

Since we have assumed in this section that we start from a single site operator, as we did in \cref{sec:master-eq}, we have that the average probabilities $\expval{a_{\Sc}^2(t)}$ only depend on the total weight $w$ and weight $w_1$ on site 1, as explained in \cref{sec:master-eq}.
Thus, we may rewrite \cref{eq:average-c} in terms of $h_t(w,w_1)$, using \cref{eq:h-def-norm}.
A similar calculation to the one leading to \cref{eq:htilde-master-eq} yields
\begin{align}
\label{eq:cc-exact-hww1}
\expval{\Cc(r,t)} =&
4\sum_{\substack{\Omega_{\Sc}\subset \{1,\cdots,N\}\\r\in\Omega_{\Sc}}}3^{\abs{\Omega_{\Sc}}-1}\frac{h_t(\abs{\Omega_{\Sc}},\abs{\Omega_{\Sc}\cap\{1\}})}{3^{\abs{\Omega_{\Sc}}}\mqty(N-1\\\abs{\Omega_{\Sc}}-\abs{\Omega_{\Sc}\cap\{1\}})}\\
=&
\frac{4}{3}\qty[\sum_{w=1}^{N-1}\mqty(N-2\\w-1)\frac{h_t(w,0)}{\mqty(N-1\\w)}+\sum_{w=2}^{N}\mqty(N-2\\w-2)\frac{h_t(w,1)}{\mqty(N-1\\w-1)}]\\
=&
\frac{4}{3(N-1)}\sum_{w=1}^{N}\qty[(w-1)h_t(w)+h_t(w,0)],
\end{align}
where $h_t(w)\equiv h_t(w,0)+h_t(w,1)$, as defined in the main text and in \cref{sec:cont-approx} [\cref{eq:h-total-def}].

Using the fact that $h_t(w)$ is normalized (i.e.\ $\sum_w h_t(w)=1$) and defining the mean weight $\expval{w(t)}= \sum_w wh_t(w)$, we get 
\begin{equation}
\expval{\Cc(r,t)} = 
\frac{4}{3}\frac{\expval{w(t)}-1}{N-1}+\frac{4}{3(N-1)}\sum_{w=1}^{N}h_t(w,0).
\end{equation}
By the normalization of the probability distribution, we further know that $\sum_{w=1}^N h_t(w,0)<1$. Hence, the second term in the equation above scales as $O(1/N)$ and is therefore negligible for large $N$.
Thus, in the limit of large $N$ we have 
\begin{dmath}
	\expval{\Cc(r,t)} =
	\frac{4}{3}\frac{\expval{w(t)}}{N} +O(1/N).
\end{dmath}
\section{Additional details on the time-evolution of $h(w,w_1)$}
\label{sec:dynamics}
In this section, we provide additional numerical and analytical details regarding the probability weight distribution.

\begin{figure}[htb]
	\centering
	\includegraphics[width=\textwidth]{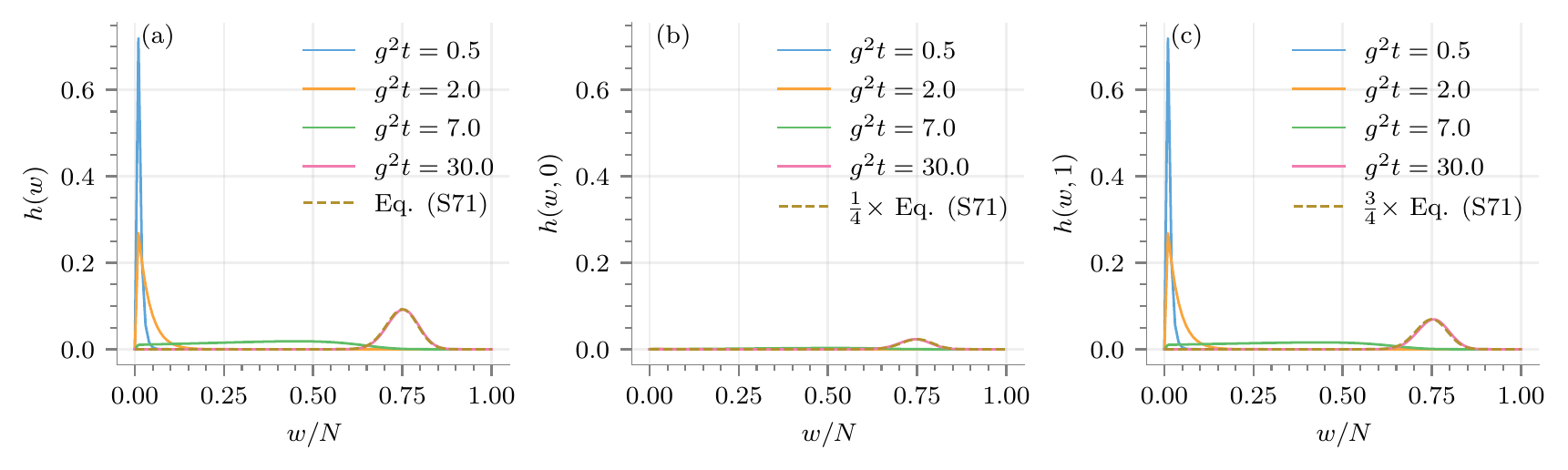}
	\caption{(a) Snapshots of the numerically computed total probability weight distribution $h(w) = h(w,0) + h(w,1)$ for $g=0.1$ and $N=100$, together with the analytical expression of the steady-state from \cref{eq:h_asymp}, which essentially agrees with the $g^2 t = 30$ numerics.  (b) The same plot for $h(w, w_1 = 0)$. While $h(w,0) \approx 0$ for early and intermediate times, the numerics for $g^2 t = 30$ essentially agree with the analytical prediction for the steady state. (c) The same plot for $h(w,w_1 = 1)$.}
	\label{fig:prob-dist}
\end{figure}

In \cref{fig:prob-dist}, we plot snapshots of $h(w)$ and $h(w,w_1=0,1)$, at different times, computed numerically using the exact master equation. 
The initial distribution starts in the $w_1=1$ sector and quickly (exponentially) expands.
At early times, during the exponential growth, the distribution is supported almost exclusively on the $w_1=1$ sector. 
At later times, when $h(w)$ is very broad in weight space and has large support on weights $w\sim O(N)$, the coupling between the two $w_1=0,1$ sectors turns on and $h(w,0)$ starts to get populated. 
Finally, $h(w)$ reaches the steady-state, which, as we show below, is, to a good approximation, a Gaussian centered at $w=3N/4$ with a width $\sim \Delta w/N \propto 1/\sqrt{N}$.
The steady-state corresponds to all strings being equally likely, and hence the Gaussian peak in $h(w,1)$ is three times as large as the one in $h(w,0)$.

\subsection{Stationary solution for $h(w)$}

At large $t$ the distribution $h(t, \phi=w/N)$ approaches a stationary solution that obeys following equation
\begin{equation}\label{eq:station_h_pde}
-\partial_\phi \left[ D_1(\phi)h(\phi) \right] + \partial^2_\phi \left[ D_2(\phi) h(\phi) \right] = 0,
\end{equation}
where
\begin{equation}
D_1(\phi) = \frac{2}{3}\phi \left(1-\frac{4 \phi}{3}\right), \qquad D_2(\phi) = \frac{\phi}{3 N} \left(1-\frac{2 \phi}{3}\right).
\end{equation}
Integrating out Eq. \ref{eq:station_h_pde} we obtain
\begin{equation}\label{eq:pde_integr}
-D_1(\phi)h(\phi) + \partial_\phi \left[ D_2(\phi) h(\phi) \right] = C.
\end{equation}
Equation (\ref{eq:pde_integr}) can be rewritten as
\begin{equation}\label{eq:pde_simpl}
\partial_\phi h(\phi) = \left(\frac{D_1(\phi)-\partial_\phi D_2(\phi)}{D_2(\phi)}\right)h(\phi) + \frac{C}{D_2(\phi)}.    
\end{equation}
Solution of (\ref{eq:pde_simpl}) is straightforward:
\begin{equation}
\begin{split}
h(\phi) &= const \times e^{J(\phi)} \int_0^\phi \frac{d\phi' e^{-J(\phi')}}{D_2(\phi')}, \\
J(\phi) &=\int d\phi \frac{D_1 - \partial_\phi D_2}{D_2} = 4N \phi-\log{\phi} + (3N-1) \log{(3-2\phi)}.
\end{split}
\end{equation}

As a result we obtain solution for $h(\phi)$ in the form:
\begin{equation}\label{eq:pde_h}
h(\phi) = const \times \frac{e^{N S(\phi)}}{(3-2\phi) \phi} \int_0^\phi d\phi' \; e^{-N S(\phi') },
\end{equation}
where  
\begin{equation}
S(\phi) = 4\phi+3\log{(3-2\phi)}.
\end{equation}

In the limit $N\to\infty$ the main contribution in the integral (\ref{eq:pde_h}) comes from the vicinity of the  boundary point $\phi=0$. Expanding  $S(\phi)$ in Taylor series in powers $\phi$: $S(\phi)\approx S(0) + 2\phi$ and substituting it inside of the integrand in Eq. (\ref{eq:pde_h})  results in
\begin{equation}\label{eq:h_asymp}
h(\phi) \sim \frac{e^{N S(\phi)}}{(3-2\phi)\phi}\left[1-e^{-2N\phi}\right].
\end{equation}
Expression Eq. (\ref{eq:h_asymp}) can be further simplified since $e^{N S(\phi)}$ is strongly peaked in the vicinity of $\phi_0=3/4$ which is the extremum of $S(\phi)$: $S(\phi)\approx S(\phi_0) + \frac{S''(\phi_0)}{2}(\phi-\phi_0)^2 + ...$, that gives
\begin{equation}
h(\phi)\sim \frac{e^{-\frac{8 N}{3}(\phi-3/4)^2}}{\phi(3-2\phi)}\left[1-e^{-2N\phi}\right].
\end{equation}

\section{Mean-weight after one step and scrambling in O(1)}
\label{sec:onestep}
In this section, we derive a simple expression for the mean-weight after a single step of the random circuit.
Here, a single step is defined as in \cref{sec:W-matrix}, i.e $U=\Uhar\Uzz\Uhar$. In doing so, we show that if the global interactions are sufficiently strong (i.e if $g'$ is independent of $N$) then a single step of the circuit is sufficient to achieve scrambling.

Starting from the initial conditions defined in \cref{eq:init-cond-full}, and using the master equation in \cref{eq:htilde-master-eq}, we find after a single step 
\begin{equation}
h_{t=1}(w,w_1) =\mathcal{R}(w,w_1,1,1).
\end{equation}
Using \cref{eq:tildeW-def,eq:W-def}, we can further simplify
\begin{equation}
h_{t=1}(w,w_1)=3^w {N-1\choose w-w_1} W(w,1,w_1)=\begin{cases}
0 \qif w_1=0,\\
\frac{1}{3}{N-1\choose w-1}\qty(\delta_{w,1}+2\qty[\cos[2](g')]^{2(N-w)}\qty[\sin[2](g')]^{2(w-1)}) \qif w_1=1.
\end{cases}
\end{equation}
The above describes the probability weight distribution after a single step, valid for arbitrary $g'$.

The mean of the above distribution can be computed exactly,
\begin{equation}
\expval{w}=\sum_{w=1}^N w h_{t=1}(w,w_1=1) = 
\frac{1}{3}+\frac{2}{3}\cos[2](g')+\frac{2}{3}N\sin[2](g').
\end{equation}
Thus, if $g'$ is independent of $N$, then $\expval{w}$ is $O(N)$ and $\expval{\Cc}$ (see \cref{sec:squared-comm-w}) is $O(1)$ after just a single step.


%